\documentstyle[prl,twocolumn,aps,epsfig]{revtex}

\begin{document}

\preprint{E895 Internal Draft}
\title{Azimuthal Dependence of Pion Interferometry at the AGS}

\author{
M.A.~Lisa$^{(1)}$,
N.~N.~Ajitanand$^{(2)}$, J.~M.~Alexander$^{(2)}$, M.~Anderson$^{(3)}$,
D.~Best$^{(4)}$, F.P.~Brady$^{(3)}$,
T.~Case$^{(4)}$, W.~Caskey$^{(3)}$, D.~Cebra$^{(3)}$, J.L.~Chance$^{(3)}$,
   P.~Chung$^{(2)}$, B.~Cole$^{(5)}$, K.~Crowe$^{(4)}$,
A.C.~Das$^{(1)}$, J.E.~Draper$^{(3)}$,
M.L.~Gilkes$^{(2)}$, S.~Gushue$^{(6,2)}$,
M.~Heffner$^{(3)}$, A.S.~Hirsch$^{(7)}$, E.L.~Hjort$^{(7)}$, L.~Huo$^{(8)}$,
M.~Justice$^{(9)}$,
M.~Kaplan$^{(10)}$, D.~Keane$^{(9)}$, J.C.~Kintner$^{(11)}$, J.~Klay$^{(3)}$, D.~Krofcheck$^{(12)}$,
R.~A.~Lacey$^{(2)}$, J.~Lauret$^{(2)}$, H.~Liu$^{(9)}$, Y.M.~Liu$^{(8)}$,
R.~McGrath$^{(2)}$, Z.~Milosevich$^{(10)}$,
G.~Odyniec$^{(4)}$, D.L.~Olson$^{(4)}$,
S.Y.~Panitkin$^{(9)}$, C.~Pinkenburg$^{(2)}$, N.T.~Porile$^{(7)}$,
G.~Rai$^{(4)}$, H.G.~Ritter$^{(4)}$, J.L.~Romero$^{(3)}$,
R.~Scharenberg$^{(7)}$, L.~Schroeder$^{(4)}$, B.~Srivastava$^{(7)}$,
  N.T.B.~Stone$^{(4)}$, T.J.M.~Symons$^{(4)}$,
R.~Wells$^{(1)}$, 
J.~Whitfield$^{(10)}$, T.~Wienold$^{(4)}$, R.~Witt$^{(9)}$, L.~Wood$^{(3)}$, and W.N.~Zhang$^{(8)}$
\\(E895 Collaboration)
}
\address{
$^{(1)}$The Ohio State University, Columbus, Ohio 43210\\
$^{(2)}$Depts. of Chemistry and Physics, SUNY Stony Brook, New York 11794-3400 \\
$^{(3)}$University of California, Davis, California, 95616 \\
$^{(4)}$Lawrence Berkeley National Laboratory, Berkeley, California, 94720\\
$^{(5)}$Columbia University, New York, New York 10027 \\
$^{(6)}$Brookhaven National Laboratory, Upton, New York 11973 \\
$^{(7)}$Purdue University, West Lafayette, Indiana, 47907-1396 \\
$^{(8)}$Harbin Institute of Technology, Harbin, 150001 P.~R. China \\
$^{(9)}$Kent State University, Kent, Ohio 44242 \\
$^{(10)}$Carnegie Mellon University, Pittsburgh, Pennsylvania 15213\\
$^{(11)}$St. Mary's College, Moraga, California  94575 \\
$^{(12)}$University of Auckland, Auckland, New Zealand \\
 }

\date{\today}
\maketitle
\begin{abstract}
Two-pion correlation functions, measured as a function of azimuthal emission
angle with respect to the reaction plane, provide novel information
on the anisotropic shape and orientation of the pion-emitting zone
formed in heavy ion collisions.
We present the first experimental determination of this information,
for semi-central Au+Au collisions at 2-6 AGeV.
The source extension perpendicular to the reaction plane is greater than
the extension in the plane, and
tilt of the pion source in
coordinate space is found to be opposite its tilt in momentum space.
\end{abstract}
\vspace{0.5cm}


Nuclear collisions at finite impact parameter are intrinsically anisotropic
with respect to the azimuthal angle about the beam direction.  
For nearly two decades, a growing community has studied the details
of the momentum-space anisotropy of emitted particles (directed and elliptic flow)
at all collision energies~\cite{flow-reviews}.
Theoretical studies~\cite{nutcracker,heiselberg,frankfurt-hydro,W98,LHWprl,VoloshinHBTflow}
suggest that {\it coordinate-space} anisotropies are equally interesting.
Recently, a formalism was proposed to use azimuthally-sensitive intensity interferometry,
correlated event-by-event with the reaction plane,
to extract this information experimentally~\cite{W98,LHWprl}.
In this Letter, we present the first experimental measurement of the full
coordinate-space anisotropies of a hot nuclear source, from pions emitted from Au+Au collisions
at 2-6 AGeV.

Two-particle interferometry (HBT) is the most direct probe of 
the space-time structure of the hot system formed in high energy heavy ion 
collisions~\cite{HBTreview,WH99}, and, along with flow, the past decades have also seen
extensive pion HBT measurements map out
the space-time geometry and dynamics of heavy ion collisions 
from Bevalac ($\sim$1 AGeV) energies~\cite{HBTBevalac},
at the AGS (2-11 AGeV)~\cite{E895HBTprl,lisa-qm99,E877,E877qm95},
and to the highest available energy ($\sim$160 AGeV) at the CERN SPS~\cite{NA44,NA49HBT-CRIS98}.
With the exception of some recent preliminary studies~\cite{lisa-qm99,E877qm95,ParkCity99},
all experimental studies to date have assumed cylindrical symmetry of the
source about the beam axis, strictly valid only for $b=0$.



The E895 Collaboration at the AGS used a large acceptance detector to measure charged particles from Au+Au
collisions.
For every event, the charged particle multiplicity is used to estimate the 
the magnitude of the impact parameter vector $|{\bf b}|$, with an uncertainty $\sim$0.5~fm.
For all data presented here, the estimated impact parameter range
is $|{\bf b}|$=4-8~fm.
The direction of ${\bf b}$ is determined from the azimuthally anisotropic momentum distribution of
protons and light nuclear fragments (not pions),
with an estimated uncertainty of $\sim 20-35^\circ$.
Further details are available elsewhere~\cite{E895elliptic,E895directed}.

Experimental details, such as correction for momentum resolution and particle identification,
of the (azimuth-integrated) HBT analysis have been published~\cite{E895HBTprl,lisa-qm99}.
Here, we discuss points particular to the azimuthally-sensitive analysis.

As in ``standard'' HBT analyses, 
correlation functions are constructed by dividing the two-pion
yield as a function of relative momentum ${\bf q}={\bf p_1} - {\bf p_2}$, 
by a background generated by mixing pions of
the current event with those of previous events~\cite{kopylov}.  
As is well-known, single-particle acceptance/efficiency and phase-space effects
cancel out with this ``event-mixing'' technique.
However, there is a special consideration here.
In the present analysis, we generate correlation functions with
selections on the pair angle with respect to the reaction plane
$\phi=\angle({\bf K_\perp},{\bf b})$, where ${\bf K_\perp}=({\bf p_1} + {\bf p_2})_\perp$
is the total momentum of the pair perpendicular to the beam~\cite{pair-cut}.  Thus, we only mix events
which have similar (within $5^\circ$) directions of reconstructed ${\bf b}$.

The correlation function was binned according to
${\bf q}$-components in the Bertsch-Pratt (``out-side-long'') decomposition as measured in
the Au+Au c.m. frame~\cite{BP}.  
In this Letter, we use the subscripts ``o, s, l'' to stand for ``out, side, long.''
Here, $q_{l}$ is the component parallel to the beam,
$q_o$ is the component parallel to ${\bf K}$, and $q_s$ is perpendicular to $q_l$ and $q_o$.
Particles in a pair were ordered such that $q_l > 0$; the signs of the other two components were
retained.  

\begin{figure}
\vspace*{-0.9cm}
\begin{center}
\epsfig{file=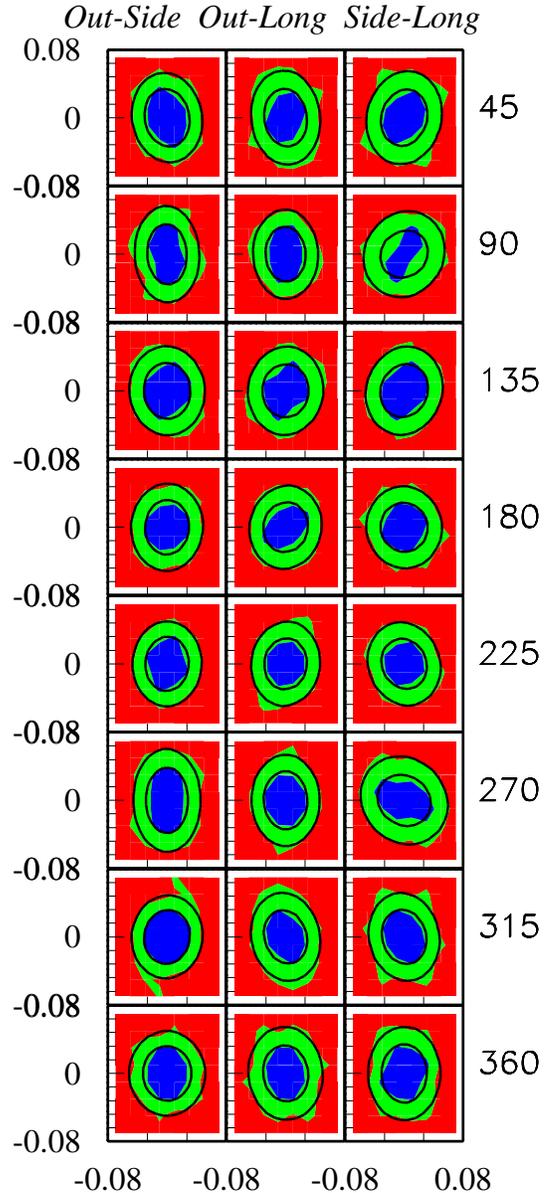,width=8cm}
\end{center}
\vspace*{-0.2cm}
\caption{
Two-dimensional projections in $q_{o}-q_{s}$ (left column),
$q_{o}-q_{l}$ (center column), and $q_{s}-q_{l}$ (right column),
of the correlation functions measured for midrapidity
pions from semi-central Au+Au collisions at 4 AGeV.  For each projection,
the unplotted component is integrated over $\pm30$ MeV/c.
Indicated by the labels on the right, projections are shown
for emission angles with respect to the reaction plane
$\phi=45^\circ\pm22.5^\circ$ (top row) to $\phi=360^\circ\pm22.5^\circ$ (bottom row).
Solid lines show projections of fits with Equation~\ref{eq:extended-BP}.
\label{fig:2dcf}}
\end{figure}

Figure~\ref{fig:2dcf} shows two-dimensional projections of the correlation function for 
4 AGeV Au+Au semicentral collisions, measured in eight $45^\circ$-wide bins in $\phi$.
The average rapidity and $p_T$  of pions in low-$q$ pairs was 1.2 (midrapidity) and 110 MeV/c, respectively,
and did not vary with $\phi$-bin.

Especially in the $q_o-q_l$ and $q_s-q_l$ projections, the correlation functions display
a distinct ``tilt,'' which evolves with the emission angle $\phi$.
As discussed below, this tilt arises from fundamental geometry~\cite{LHWprl}, in contrast to a similar tilt
observed~\cite{NA49HBT-CRIS98} in some  $q_o-q_l$ projections measured far from midrapidity
for azimuthally-integrated HBT, which arises from dynamic (longitudinal flow) effects~\cite{chapman}.
Aside from these dynamical effects, there are no 2-dimensional tilts in azimuthally-integrated HBT,
and only one-dimensional projections are typically shown~\cite{lisa-qm99}.


Using a maximum-likelihood technique, we fit the correlation functions for each of the
$\phi$ bins with the standard Gaussian parameterization~\cite{WH99}
\begin{equation}
  \label{eq:extended-BP}
    C({\bf q},\phi) =
    1 + \lambda(\phi)\, 
    \exp\Bigl[- \sum_{i,j=o,s,l} q_i q_j R_{ij}^2(\phi) \Bigr]
    \,.
\end{equation}
In contrast to azimuthally-integrated analyses, all six radius parameters are relevant
here~\cite{W98,LHWprl,WH99}.

Two-dimensional projections of the fits, weighted according to the mixed-event background,
are shown as contours in Figure~\ref{fig:2dcf}.
Since the relative sign of any component of ${\bf q} = {\bf p_1} - {\bf p_2}$
is arbitrary (while the relative sign between components is not arbitrary),
the two-dimensional projections in Figure~1 have been reflected for clarity.
However, the fits were performed in the full three dimensions, with each
pion pair being weighted only once.

\begin{figure}
\vspace*{-0.9cm}
\begin{center}
\epsfig{file=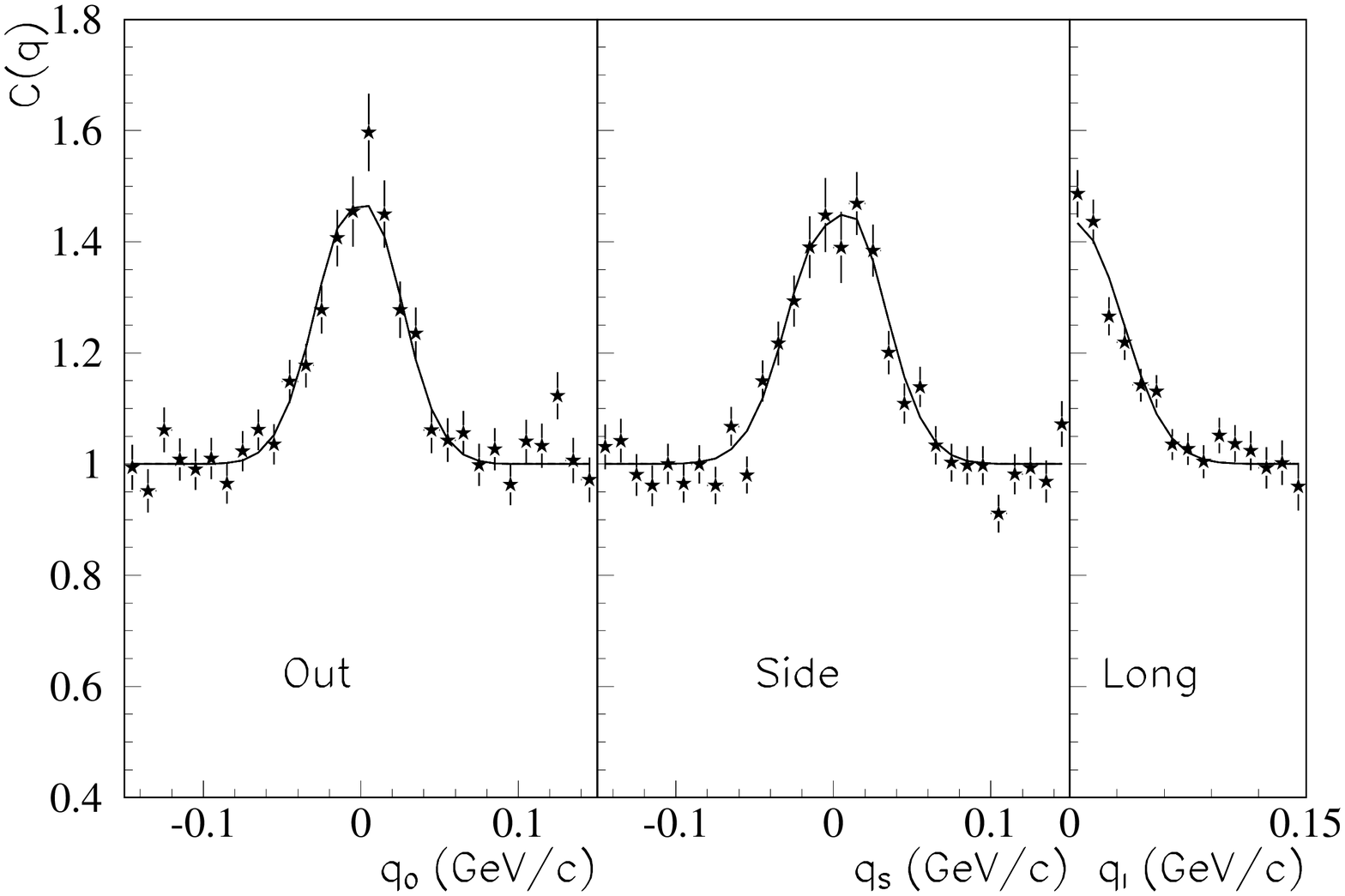,width=9cm}
\end{center}
\vspace*{-0.2cm}
\caption{
One-dimensional projections, in $q_o$, $q_s$, and $q_l$, of the correlation function
for $\phi=45^\circ\pm22.5^\circ$ (top row of Figure~\ref{fig:2dcf}).  For each projection,
the unplotted $q$ components are integrated over $\pm30$ MeV/c.  Note that in the
present analysis,  $q_l$ is defined positive, while $q_o$ and $q_s$ have a meaningful
signs.  Solid lines show projections of fits with Equation~\ref{eq:extended-BP}.}
\label{fig:1dcf}
\end{figure}

Although the full structure of the correlation function is only
visible in multi-dimensional projections such as those in
Figure~\ref{fig:2dcf}, one-dimensional projections are commonly
shown for azimuthally-integrated HBT analyses, to provide a feeling
for the quality of the data and fit.  For reference, one-dimensional
projections of the correlation function and fits,
for one $\phi$ slice, are shown in Figure~\ref{fig:1dcf}.

\begin{figure}
\vspace*{-0.9cm}
\begin{center}
\epsfig{file=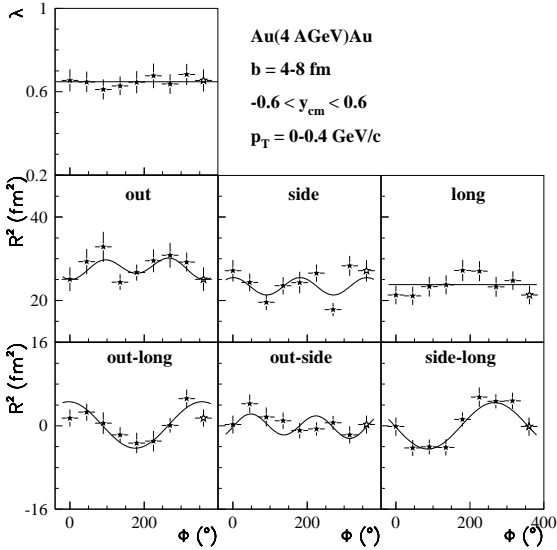,width=8cm}
\end{center}
\vspace*{-0.2cm}
\caption{
Filled stars show the fit parameters from Equation~\ref{eq:extended-BP}, resulting from
fits to the correlation functions of Figure~\ref{fig:2dcf}.
The values at $\phi=0^\circ$ are redisplayed as open stars at $\phi=360^\circ$.
The line in the $\lambda$ panel represents the average value of $\lambda$.  Lines in the 
other panels represent the fit to the HBT radii (stars) with the Equation~\ref{eq:Wiedemann-fit}.
The vertical scale is linear in all panels.
\label{fig:4gev-radii}}
\end{figure}

The seven fit parameters $R_{ij}$ and $\lambda$ are plotted as a function of $\phi$ in
Figure~\ref{fig:4gev-radii}.
As is well-known~\cite{HBTreview,WH99}, the HBT ``radii'' parameters are in general not
strictly the radii of the emitting source, but have a geometrical interpretation as
lengths of homogeneity of the emitting region.  For realistic sources, various
interpretations of $\lambda$ have been proposed (see e.g.~\cite{nickerson98,hardtke-voloshin2000}).
At AGS energies, the observed strong energy dependence of $\lambda$ was well reproduced~\cite{E895HBTprl}
by detailed simulations using the RQMD (v2.3) transport model~\cite{RQMD}; there,
the decrease of $\lambda$ with energy was attributed to increased production of long-lived
$\pi^-$-emitting particles at higher energy (see also~\cite{sullivan94}).  In this scenario
then, the HBT radii characterize a Gaussian ``core'' of directly-produced pions,
with $\lambda < 1$ resulting from a well-separated ``halo'' of pions emitted from long-lived
particles~\cite{csorgo-core-halo}.  In this framework, we now discuss the $\phi$-dependence
of the HBT parameters, and how to extract the underlying geometry of the ``core'' of the source.

The $\phi$-indepdence of $\lambda$ indicates that the fraction of $\pi^-$ from long-lived
resonances is independent of emission angle with respect to the reaction plane.  This seems
reasonable, given the relatively weak anisotropic flow at these energies~\cite{E895elliptic},
as well as the randomizing effect on the pion from decay of the parent.

$R_{o}$ and $R_{s}$ show significant
equal and opposite
second-order oscillations in $\phi$, consistent with the simple picture of a transverse
profile of the pion-emitting region that reflects the ``almond-shaped'' overlap region
(with a larger spatial extent perpendicular to ${\bf b}$ than parallel to ${\bf b}$)
 between the target and projectile
spheres~\cite{heiselberg,lisa-qm99}.

Note that the ``cross-term'' radii ($R_{ol}^2$, $R_{os}^2$, $R_{sl}^2$) quantify the tilt
observed in the 2-dimensional correlation function projections of each $\phi$-bin of Figure~\ref{fig:2dcf}.
$R_{ol}^2$ and $R_{sl}^2$ display equal-magnitude first-order oscillations, consistent with pion emission
from an ellipsoidal source which is tilted in coordinate space in the reaction plane, and away from the beam
axis~\cite{LHWprl}.

In general, the six HBT radii are related to the spatiotemporal structure of the emitting source via
the equations~\cite{W98,LHWprl}
  \begin{eqnarray}
    && R_s^2 = S_{11} \sin^2\phi
                  + S_{22} \cos^2\phi
                  - S_{12} \sin 2\phi \, ,
                      \nonumber \\
    && R_o^2 = S_{11} \cos^2\phi 
                  + S_{22} \sin^2\phi
                  + S_{12} \sin 2\phi
   \nonumber \\
    && \qquad  \quad 
                     - 2\beta_\perp S_{01} \cos\phi 
                     - 2\beta_\perp S_{02} \sin\phi 
                     + \beta_\perp^2 S_{00} \, ,
   \nonumber \\
    && R_{l}^2 = S_{33} -2 \beta_l S_{03} +
                       \beta_l^2 S_{00} \, ,
   \nonumber \\
    && R_{os}^2 = 
                  S_{12} \cos 2\phi 
                  + \textstyle{1\over 2} \left(S_{22}-S_{11}\right)
                  \sin 2\phi 
      \label{eq:Wiedemann-fit} \\
    && \qquad  \quad \ \ 
                     + \beta_\perp S_{01} \sin\phi
                     - \beta_\perp S_{02} \cos\phi \, ,
   \nonumber \\
    && R_{ol}^2 = 
                  \left( S_{13} - \beta_l S_{01}\right) \cos\phi
                  - \beta_\perp S_{03} 
   \nonumber \\
    &&\qquad \quad \ 
                  + \left( S_{23} - \beta_l S_{02}\right) \sin\phi
                  + \beta_l\beta_\perp S_{00}\, ,
   \nonumber \\
    && R_{sl}^2 = 
                  \left(S_{23} - \beta_l S_{02}\right) \cos\phi
                  - \left( S_{13} - \beta_l S_{01}\right) \sin\phi\, . \nonumber
  \end{eqnarray}
where $\beta_\perp$ and $\beta_l$ are the average pair velocities in the transverse and
longitudinal direction, and the spatial correlation tensor $S_{\mu\nu}$ is given by
\begin{equation}
  S_{\mu\nu} = \langle \tilde x_\mu \tilde x_\nu \rangle\, ,
  \quad \tilde{x}_\mu = x_\mu - \bar x_\mu\, ,
  \quad (\mu,\nu=0,1,2,3)
  \label{eq:Smunu}
\end{equation}
where the brackets $\langle \, \rangle$ indicate an average over the emitting source.
In Equations~\ref{eq:Wiedemann-fit}, $\mu=3$ refers to the beam direction, and $\mu=1$
is parallel to ${\bf b}$.


Simultaneously fitting the six $R_{ij}^2(\phi)$ of Figure~\ref{fig:4gev-radii} with Equations~\ref{eq:Wiedemann-fit}
(treating the $S_{\mu\nu}$ as 10 $\phi$-independent fit parameters) allows
extraction of the full spatial correlation tensor
\begin{displaymath}
 S = \left( \begin{array}{rrrr}

   13.4  \pm 3.5  &   0.7  \pm 0.9  &   0.2  \pm 1.0  &  0.0  \pm 1.1  \\
    0.7  \pm 0.9  &  21.3  \pm 0.9  &  -0.1  \pm 0.6  &   4.5  \pm 0.6  \\
    0.2  \pm 1.0  &  -0.1  \pm 0.6  &  25.5  \pm 1.0  &  -0.3  \pm 0.6  \\
    0.0  \pm 1.1  &  4.5   \pm 0.6  &  -0.3  \pm 0.6  &  22.8  \pm 0.9  
	\end{array} \right) \nonumber
\end{displaymath}
%
where all units are in fm$^2$ and $c=1$.
The fits are shown by solid lines in Figure~\ref{fig:4gev-radii}.

The only significantly non-vanishing tensor elements are the diagonal ones $S_{\mu\mu}$ and $S_{13}$.
This fact, along with observations of identical oscillation amplitudes in $R_{o}^2$ and $R_s^2$,
and in $R_{ol}^2$ and $R_{sl}^2$, support the assumption of $\phi$-independence of $S_{\mu\nu}$
in the fit~\cite{W98,LHWprl}.


$S_{13}$ and $(S_{22}-S_{11})$ quantify, respectively, first and second harmonic
oscillations of the HBT radii with respect to the measured reaction plane.  
However, due to finite particle multiplicity,
the measured reaction plane differs statistically from the true one
by some angle $\Delta\phi$~\cite{E895elliptic,resolution-correction-theory}.
This reaction plane dispersion results in an apparent reduction in both 
$S_{13}$ and $(S_{22}-S_{11})$
(but does not affect quantities $S_{00}$, $S_{33}$, and ($S_{11}+S_{22}$)).  According
to~\cite{W98,resolution-correction-theory,voloshin-private}
\begin{eqnarray}
S_{13,m} = S_{13} \cdot \langle\cos(\Delta\phi)\rangle                      \\
(S_{22}-S_{11})_m =  (S_{22}-S_{11}) \cdot \langle\cos(2\Delta\phi)\rangle ,
\end{eqnarray}
where $S_{13,m}$ is the observed value, and $S_{13}$ is the ``true'' value.
The correction factors obtained by
following the procedure  outlined in Refs.~\cite{W98,E895elliptic,E895directed,resolution-correction-theory}
are listed in Table~\ref{tab:corrections}.
Hence, for the 4 AGeV data, the reaction plane-dispersion corrected values are
$S_{11}=19.8\pm1.2$, fm$^2$, $S_{22}=27.0\pm1.4$ fm$^2$, and $S_{13}=5.2\pm0.7$ fm$^2$.

The spatial tilt of the emission ellipsoid in coordinate space is given by~\cite{LHWprl}
\begin{equation}
\theta_s = \frac{1}{2} \tan^{-1}
       \left( \frac{2S_{13}}{S_{33}-S_{11}}\right)
= 37^o \pm 4^o .
\end{equation}
%
Rotating the spatial correlation tensor (corrected for reaction-plane dispersions) by this
angle returns the diagonal tensor
\begin{eqnarray*}
R^\dagger_y(\theta_s)\cdot S \cdot R_y(\theta_s) =~~~~~~~~~~~~~~~~~~~~~~~~~~  \nonumber \\
\left( \begin{array}{rrrr}
   13.4  \pm 3.5  &   0.5  \pm 1.0  &   0.2  \pm 1.0  &  0.4  \pm 1.0  \\
    0.5  \pm 1.0  &  15.9  \pm 1.0  &   0.1  \pm 0.6  &   0.0  \pm 0.9  \\
    0.2  \pm 1.0  &   0.1  \pm 0.6  &  27.0  \pm 1.4  &  -0.3  \pm 0.6  \\
    0.4  \pm 1.0  &  0.0   \pm 0.7  &  -0.3  \pm 0.6  &  26.8  \pm 0.8  

	\end{array} \right) \nonumber
\end{eqnarray*}
whose diagonal elements are the squared lengths of homogeneity ($S_{\mu\mu}=\sigma_\mu^2$)
in the t, x, y, z directions.

\begin{figure}
\vspace*{-0.9cm}
\begin{center}
\epsfig{file=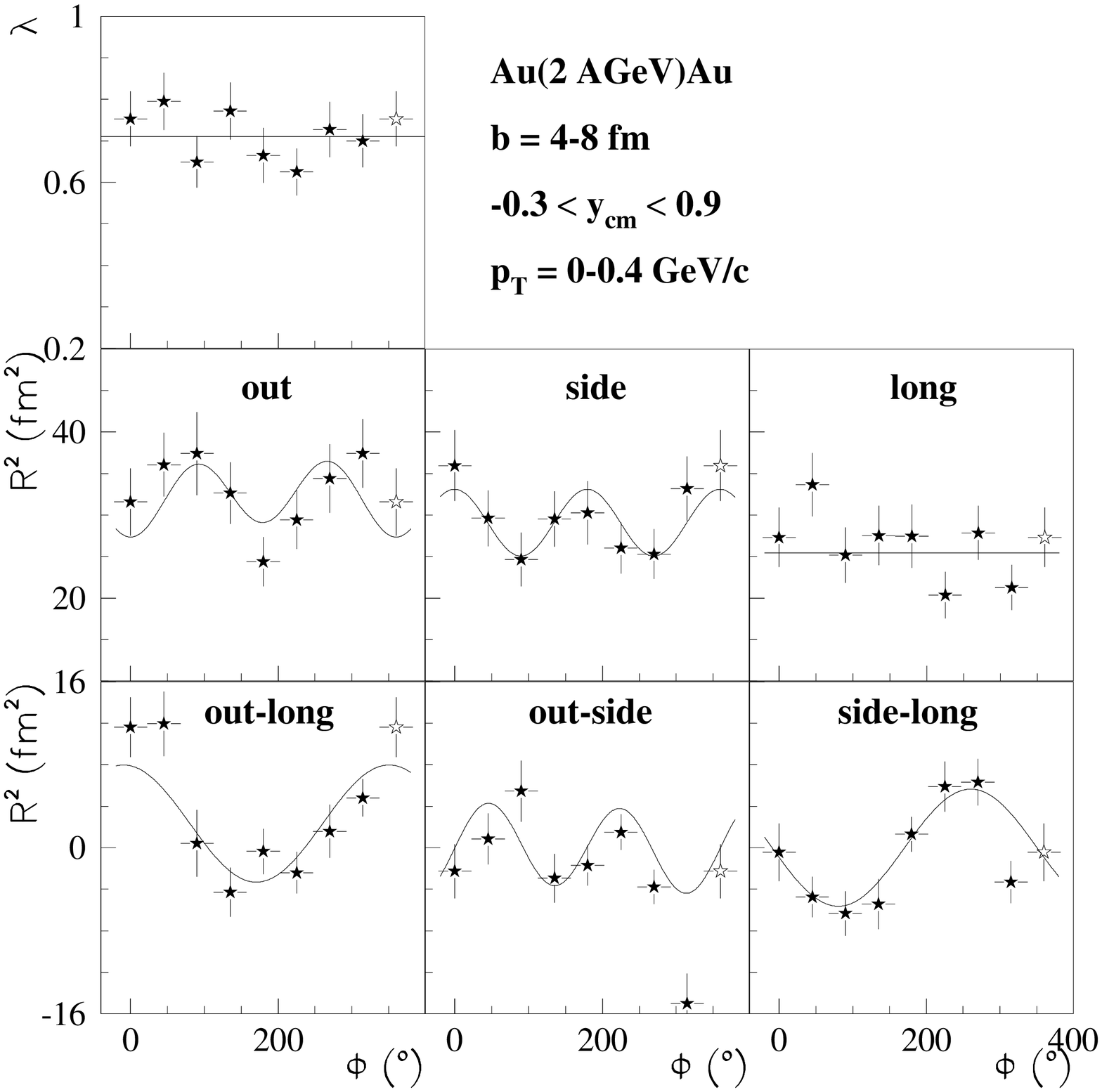,width=8cm}
\end{center}
\vspace*{-0.2cm}
\caption{
Same as Figure~\ref{fig:4gev-radii}, but for 2 AGeV collisions.
\label{fig:2gev-radii}}
\end{figure}

\begin{figure}
\vspace*{-0.9cm}
\begin{center}
\epsfig{file=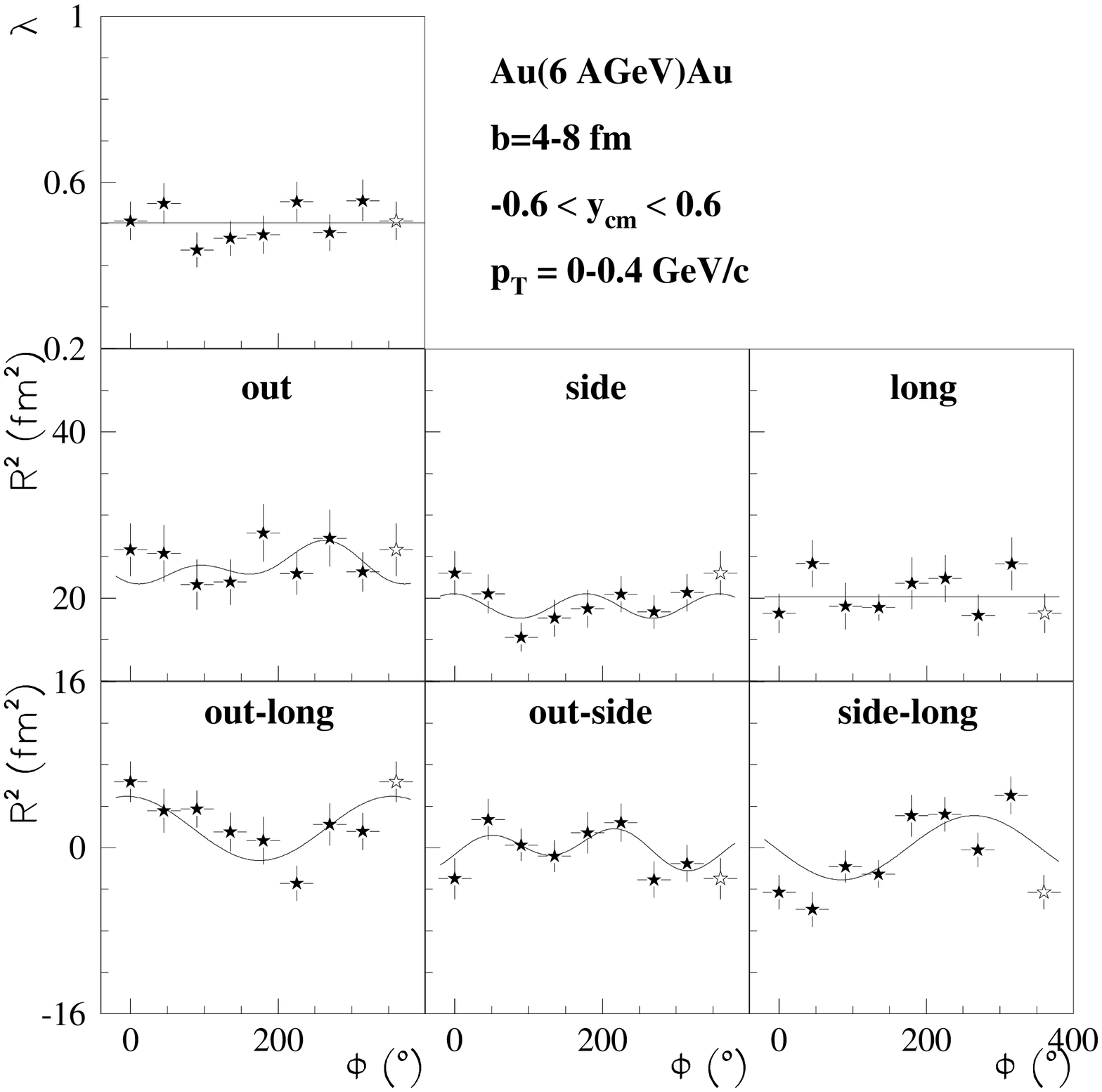,width=8cm}
\end{center}
\vspace*{-0.2cm}
\caption{
Same as Figure~\ref{fig:4gev-radii}, but for 6 AGeV collisions.
\label{fig:6gev-radii}}
\end{figure}

Similar results are obtained at other energies.  Figures~\ref{fig:2gev-radii} and ~\ref{fig:6gev-radii} show the HBT parameters
obtained at 2 and 6~AGeV~\cite{footnote-2GeV,footnote-6GeV}.
Especially at the highest energy, degraded reaction-plane resolution 
significantly reduces the second-order oscillations in the transverse radii.
Considered in isolation, $R_o^2$ in Figure~\ref{fig:6gev-radii} may be fit by a constant ($\phi$-independent)
value as well as anything else.  However, the simultaneous fit of all $R^2(\phi)$
allows relatively clean extraction of the source shape parameters.
Reaction plane dispersion corrections
are applied
 to all data,
but the effects are small.
For the worst case (6 AGeV dataset), the corrections result in a decrease (increase) in $\sigma_x$ ($\sigma_y$)
of 0.3 fm, and 5$^\circ$ reduction in $\theta_s$; $\sigma_z$ and $\sigma_t$ are unchanged.

Table~\ref{tab:shapes} summarizes
the inferred pion source shapes for our three energies.
At all energies, our results indicate a pion freezeout distribution as an ellipsoid
whose major axis in the reaction plane is tilted with respect to the beam in the positive direction
(i.e. in the direction of ${\bf b}$),
and whose transverse axis perpendicular to the reaction plane is longer than the
axis in the reaction plane (the ``almond'' shape referred to above).  The extension in the temporal direction is consistent with
observations at high energy~\cite{E895HBTprl,E877,NA44,NA49HBT-CRIS98}.

The observed transverse ``almond'' shape is reminiscent of the entrance channel geometry.
In the simplest picture, pions are emitted from the overlap region of the two 
spherical Au nuclei.  For impact parameter $b=$ 4~fm (8~fm), the spatial RMS of the
overlap region is 2.2~fm (1.3~fm) parallel to ${\bf b}$, and 2.9~fm (2.4~fm) perpendicular~\cite{root5-footnote}.
While the linear scale of the freezeout distribution is roughly twice these estimates (see Table~\ref{tab:shapes}),
indicating significant expansion, the aspect ratio $\sigma_y/\sigma_x \approx 1.3-1.4$, is in the range ($1.3-1.9$)
expected from this naive picture.

On the other hand, the large positive tilt is a geometric feature of the collision dynamics~\cite{LHWprl}.
Studies with the RQMD model indicate that most of the low-$p_T$ pions arise from
$\Delta$ decay, so it is not surprising that the tilted freeze-out distribution resembles the baryonic
distribution in coordinate space calculated in transport codes~\cite{frankfurt-hydro}.


Experimental access to this level of geometric detail on the freezeout distribution
is unprecedented, and represents an exciting new opportunity to study the dynamical response
of hot nuclear matter to compression.  A detailed theoretical discussion is beyond the scope
of this Letter, but we note that an identical analysis on pions generated by the RQMD model
displays considerable sensitivity on the dynamical effect of the nuclear meanfield; RQMD values are listed
in Table~\ref{tab:shapes}.
Although the spatial scale is underpredicted at the lower energies and the temporal scale overpredicted
(noted already for central collisions~\cite{E895HBTprl}),
qualitatively, the model reproduces the ``almond'' shape and large positive
tilt angles remarkably well.  Since the model better describes proton directed flow when the mean field is included
in the calculation~\cite{E895directed},
it is interesting to note that the tilt angles
(the spatial counterpart of proton directed flow~\cite{LHWprl}) reproduce observation better when the mean field is off.

We stress that these coordinate-space anisotropies represent new information, independent of
momentum-space anisotropies (directed and elliptical flow).
Momentum-space tilt angles (flow angles~\cite{FlowAngle}) at these energies are only a few
degrees~\cite{E895directed}, and, indeed,
the directed flow (momentum-space tilt) of pions at these energies is in the {\it negative}
direction~\cite{EOS_pionflow,caskey_thesis,LHWprl}, opposite the coordinate-space tilt $\theta_s$.
Experimental information on the interplay between coordinate and momentum space anisotropies
should help resolve theoretical issues, such as the coexistence of flow and antiflow components
at these energies~\cite{frankfurt-hydro}.

Further studies of the elliptical transverse shape of the source at higher energy should prove quite interesting.
It is known, for example, that the elliptic flow of nuclear matter (protons and pions)  in {\it momentum space} changes
sign at AGS energies~\cite{caskey_thesis,E895elliptic,FOPIpionv2,E877pionv2}, 
from negative (more momentum out-of-plane) at low 
collision energy to positive.
The coordinate-space analogue of negative elliptic flow is the ``almond'' shape we observe.  
The degree to which this shape follows the momentum space and evolves to an in-plane shape at high energies,
may provide valuable information on the detailed nature and cause of elliptic flow~\cite{dani_v2};
RQMD predicts~\cite{YangSnellingsXu}
that the ``almond'' shape is retained even at RHIC energies.

In conclusion, we have presented the first full measurement of the azimuthal dependence of pion interferometry.
For semi-peripheral Au+Au collisions at 2-6 AGeV, the spatial correlation tensor, $S_{\mu\nu}$, 
extracted from the $\phi$-dependences of the HBT radius parameters, reveals an ellipsoidal pion
emission region with an ``almond'' transverse profile and which is strongly tilted in the reaction plane
away from the beam axis.  Consistency relations indicate that the extracted geometry is negligibly affected
by possible transverse space-momentum correlations.  The RQMD transport model reproduces the qualitative
features of the data quite well, and reveals a dependence of the coordinate-space anisotropies
on the action of the nuclear meanfield.
Through application of this new type of analysis at other energies and careful theoretical comparisons,
it is hoped that fresh insight on the dynamics of non-central heavy ion collisions may be gained.

M.A.L. thanks Drs. U. Heinz, A. Poskanzer, S. Voloshin, and U. Wiedemann
for important suggestions and discussions, and Drs. S. Pratt and H. Sorge for the
use of their computer codes RQMD and CRAB.

This work supported by the U.S. Department of Energy under contracts DE-AC03-76SF00098 and
DE-AC02-98CH10886, and grants DE-FG02-89ER40531, DE-FG02-88ER40408, DE-FG02-87ER40324 and 
DE-FG02-87ER40331; by the U.S. National Science Foundation under grants PHY-98-04672, PHY-97-22653,
PHY-96-01271, PHY-96-05207, and INT-92-25096; by the University of Auckland Research Committee, 
NZ/USA Cooperative Science Programme CSP 95/33; and by the National
Natural Science Foundation of P.R. China under grant number 19875012.

\begin{table}
\begin{tabular}{|c|c|c|} \hline
   E (AGeV) & $\langle\cos(\Delta\phi)\rangle$  & $\langle\cos(2\Delta\phi)\rangle$  \\ \hline
    2       & 0.940                             & 0.787                              \\ \hline
    4       & 0.853                             & 0.584                              \\ \hline
    6       & 0.724                             & 0.384                              \\ \hline
\end{tabular}
\caption{First and second harmonic reaction plane dispersion correction factors for the three datasets discussed.}
\label{tab:corrections}
\end{table}

\begin{table}
\begin{tabular}{|c|c|c|c|c|c|} \hline
   E (AGeV) & $\sigma_t$ (fm/c) & $\sigma_x$ (fm) & $\sigma_y$ (fm) & $\sigma_z$ (fm) & $\theta_s$ ($^\circ$)   \\ \hline
      2 Data    &  3.5 $\pm$ 1.0    &  4.2 $\pm$ 0.2  &  5.8 $\pm$ 0.2  &  5.4 $\pm$ 0.1  &   47 $\pm$ 5  \\
      RQMD cs   &  6.6 $\pm$ 0.2    &  2.9 $\pm$ 0.1  &  4.4 $\pm$ 0.1  &  4.4 $\pm$ 0.1  &   49 $\pm$ 2  \\
      RQMD mf   &  5.4 $\pm$ 0.2    &  3.1 $\pm$ 0.1  &  5.0 $\pm$ 0.1  &  4.6 $\pm$ 0.1  &   64 $\pm$ 2  \\ \hline
      4 Data    &  3.7 $\pm$ 0.5    &  4.0 $\pm$ 0.1  &  5.2 $\pm$ 0.1  &  5.2 $\pm$ 0.1  &   37 $\pm$ 4  \\
      RQMD cs   &  6.4 $\pm$ 0.2    &  3.3 $\pm$ 0.1  &  4.2 $\pm$ 0.1  &  4.7 $\pm$ 0.1  &   33 $\pm$ 3  \\
      RQMD mf   &  4.6 $\pm$ 0.3    &  3.5 $\pm$ 0.1  &  4.7 $\pm$ 0.1  &  4.7 $\pm$ 0.1  &   45 $\pm$ 3  \\ \hline
      6 Data    &  3.8 $\pm$ 0.5    &  3.5 $\pm$ 0.2  &  4.8 $\pm$ 0.2  &  4.7 $\pm$ 0.1  &   33 $\pm$ 6  \\
      RQMD cs   &  5.9 $\pm$ 0.2    &  3.5 $\pm$ 0.1  &  4.3 $\pm$ 0.1  &  4.9 $\pm$ 0.1  &   28 $\pm$ 3  \\
      RQMD mf   &  5.0 $\pm$ 0.3    &  3.6 $\pm$ 0.1  &  4.7 $\pm$ 0.1  &  4.5 $\pm$ 0.1  &   48 $\pm$ 5  \\ \hline
\end{tabular}
\caption{Lengths of homogeneity $\sigma_\mu$ and tilt angle at each collision energy.
In addition to experimental results, predictions of the RQMD model with meanfield off (cs) and on (mf) are given.
}
\label{tab:shapes}
\end{table}


\begin{thebibliography}{9}
\bibitem{footnote-2GeV}
  For the 2 AGeV analysis, our rapidity cut is centered somewhat forward of midrapidity, for statistics
  and acceptance reasons; hence, $S_{03}$ does not vanish~\cite{chapman}, resulting in a net offset in $R_{ol}^2(\phi)$.
\bibitem{footnote-6GeV}
  In the 6~AGeV case, the fit produces an additional non-zero component: $S_{02}=1.3 \pm 1.0$.  This results in
  the slight asymmetry that is clear in the fit lines to $R_o^2$ and $R_{os}^2$.  We do not consider significant this
  barely-nonvanishing component.
\bibitem{root5-footnote}
Recall that the RMS values are significantly less than the hard-sphere radius of 7~fm
For a uniformly-filled hard sphere of radius $R$, the RMS in any one direction is $R/\sqrt 5$.
\end{thebibliography}
\end{document}